%% file: main.tex
\pdfoutput=1 
\documentclass[sigconf]{acmart}

\AtBeginDocument{%
  }

\usepackage{pifont}
\usepackage[]{mdframed}
\usepackage{graphicx}
\usepackage{cmap}

\newcounter{ziyu}
\stepcounter{ziyu}

\newcounter{asterios}
\stepcounter{asterios}

\newcommand{\para}[1]{\vspace{1mm}\textbf{#1}.}

\copyrightyear{2024}
\acmYear{2024}
\setcopyright{rightsretained}
\acmConference[CIKM '24]{Proceedings of the 33rd ACM International Conference on Information and Knowledge Management}{October 21--25, 2024}{Boise, ID, USA}
\acmBooktitle{Proceedings of the 33rd ACM International Conference on Information and Knowledge Management (CIKM '24), October 21--25, 2024, Boise, ID, USA}
\acmDOI{10.1145/3627673.3679210}
\acmISBN{979-8-4007-0436-9/24/10}

\makeatletter \gdef\@copyrightpermission{ \begin{minipage}{0.3\columnwidth} \href{https://creativecommons.org/licenses/by/4.0/}{\includegraphics[width=0.90\textwidth]{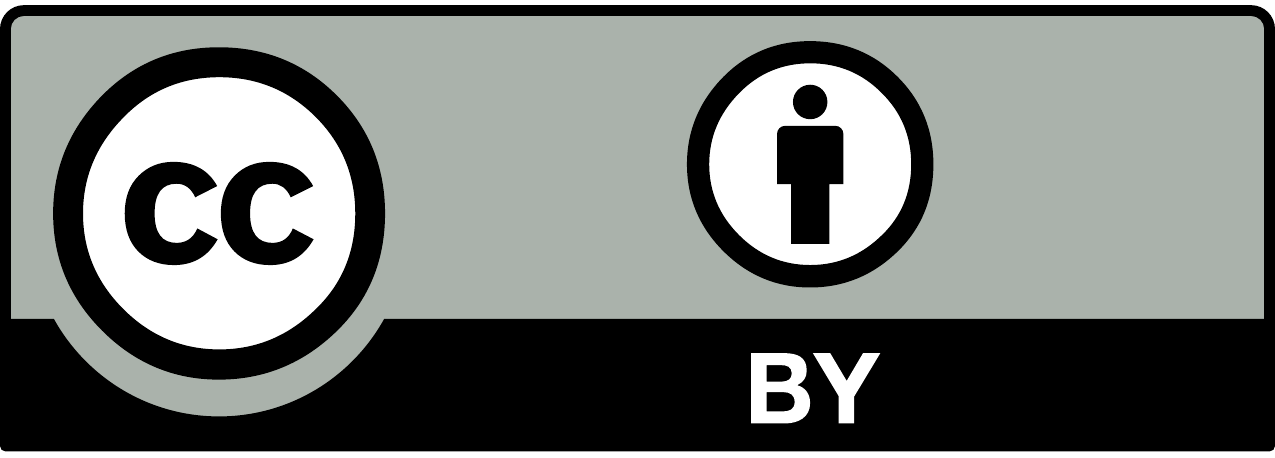}} \end{minipage}\hfill \begin{minipage}{0.7\columnwidth} \href{https://creativecommons.org/licenses/by/4.0/}{This work is licensed under a Creative Commons Attribution International 4.0 License.} \end{minipage} \vspace{5pt} } \makeatother 
\begin{document} 

\title{\texttt{LLM-PQA}: LLM-enhanced Prediction Query Answering}

\author{Ziyu Li}
\affiliation{%
\department{Department of Software Technology}
  \institution{Delft University of Technology}
  \city{Delft}
  \state{The Netherlands}
}
\email{z.li-14@tudelft.nl}

\author{Wenjie Zhao}
\affiliation{%
\department{Department of Software Technology}
  \institution{Delft University of Technology}
  \city{Delft}
  \state{The Netherlands}
}
\email{w.zhao-15@student.tudelft.nl}

\author{Asterios Katsifodimos}
\affiliation{%
\department{Department of Software Technology}
  \institution{Delft University of Technology}
  \city{Delft}
  \state{The Netherlands}
}
\email{a.katsifodimos@tudelft.nl}

\author{Rihan Hai}
\affiliation{%
    \department{Department of Software Technology}
  \institution{Delft University of Technology}
  \city{Delft}
  \state{The Netherlands}
}
\email{r.hai@tudelft.nl}

\renewcommand{\shortauthors}{Ziyu Li, Wenjie Zhao, Asterios Katsifodimos, \& Rihan Hai}

\begin{abstract}

The advent of Large Language Models (LLMs) provides an opportunity to change the way queries are processed, moving beyond the constraints of conventional SQL-based database systems.
However, using an LLM to answer a \textit{prediction} query is still challenging, since an external ML model has to be employed and inference has to be performed in order to provide an answer. 
This paper introduces \texttt{LLM-PQA}, a novel tool that addresses prediction queries formulated in natural language. 
\texttt{LLM-PQA} is the first to combine the capabilities of LLMs and retrieval-augmented mechanism for the needs of prediction queries by integrating data lakes and model zoos. 
This integration provides users with access to a vast spectrum of heterogeneous data and diverse ML models, facilitating dynamic prediction query answering. 
In addition, \texttt{LLM-PQA} can dynamically train models on demand, based on specific query requirements, ensuring reliable and relevant results even when no pre-trained model in a model zoo, available for the task. 

\end{abstract}




\maketitle

\input{sections/introduction}

\input{sections/framework}

\input{sections/demo}

\input{sections/conclusion}
\section*{Acknowledgment}
This publication is part of the project Understanding Implicit Dataset Relationships for Machine Learning (project number VI.Veni.222.439 of the research programme NWO Talent Programme Veni which is (partly) financed by the Dutch Research Council (NWO).  This work was supported by the European Union Horizon Programme (HORIZON-CL4-2022-DATA-01), under Grant 101093164 (ExtremeXP).

\bibliographystyle{ACM-Reference-Format}
\bibliography{main}

\end{document}

%% file: sections/introduction.tex
\section{Introduction}

The recent advancements of Large Language Models (LLMs) has opened up  opportunities in tackling complex language understanding tasks~\cite{kaddour2023challenges,chang2024survey}. 
These breakthroughs have inspired novel database management technologies, leading to increasing research interest in natural language to SQL ~\cite{fu2023catsql,zhang2024benchmarking}. These works allow users to formulate data retrieval queries in natural language, simplifying interactions with database management systems without requiring in-depth knowledge of SQL syntax.

Consider the scenario where a practitioner needs to know the insurance charges of a person meeting specific criteria, e.g., a 19-year-old female non-smoker, with a BMI of 27.9. 
Suppose the database contains a dataset with insurance charge information based on various features. The practitioner's requirement could be resolved by retrieving the relevant record in the database. 
However, in cases where the required data does not exist in the database, using an LLM alone could lead to unreliable results due to its tendency to generate hallucinated answers \cite{ji2023survey}.
Instead, performing ML inference with a model specifically for this task could provide a more reliable answer. This type of query, requiring ML model prediction to generate the result, is referred as prediction query~\cite{predictionquery} (or predictive query~\cite{predictivequery}). 

This scenario exemplifies and highlights the need for innovative systems that go beyond simple data retrieval, which can further incorporate advanced predictive analytics in an easy-to-use manner.
However, answering such prediction queries is challenging.
First, the user request needs to be translated into a series of actionable steps or pipelines.
With the same example, the task is to obtain a value through regression, with other information serving as input features.
Second, it would be impractical to train a new model for every query due to resource and time constraints. 
The alternative solution, finding a suitable pre-trained model, if it exists, is also a challenge task.
For instance, HuggingFace~\cite{huggingface} hosts 111 models for regression tasks, each varying widely in terms of various factors.

To address the challenges mentioned above, we propose a novel tool, \texttt{LLM-PQA}
, designed to handle prediction queries in natural language.
\texttt{LLM-PQA}\footnote{\url{https://github.com/zLizy/LLM-PQA}} integrates a data lake~\cite{hai2016constance,nargesian2019data} and a model zoo~\cite{kaggle,onnx} that serve as the sources of datasets and models.
The data lake facilitates the management of heterogeneous data across various domains. 
It provides the training data for the model training. 
While the model zoo offers a wide selection of ML models, tailored to specific analytical tasks.
To align the most suitable model with the task specified in the query, we employ a vector search strategy. In this approach, the query, models, and datasets are encoded as vectors which are served as indices. 
The model with the most similar vector to the query vector is selected as the candidate for model inference, ensuring a relevant and efficient response to the query.
Moreover, \texttt{LLM-PQA} can also deliver reliable results even when no pre-trained model is available. 
It allows ML model training ``on the spot" based on the specific requirements of the query.
The contributions of this work can be summarized as below:


\begin{itemize}
    \item \textbf{Handling prediction queries beyond standard retrieval}: \texttt{LLM-PQA} is designed to handle prediction queries, formulated in natural language. 
    This allows for a more intuitive user interaction while addressing complex analytical needs.
    \item \textbf{Matching model to query with vector search}: Another critical contribution is the innovative matching mechanism, which accurately pairs a model to a specific task given a query. 
    This tool integrates a data lake and a model zoo, which together provide access to a diverse collection of datasets and ML models, thereby facilitating precise model selection.
    \item \textbf{On-the-Spot model training capability}: \texttt{LLM-PQA} is able to  train ML models tailored to the specific need of a query, which ensures high accuracy and relevance in the responses.
\end{itemize}


%% file: sections/framework.tex
\section{Architecture}

In this section, we introduce the architecture of \texttt{LLM-PQA}, designed to answer prediction queries expressed in natural language.
We first introduce the components 
and then illustrate the workflow.


\subsection{Components}

\begin{figure}
  \centering  
  \leavevmode
  \includegraphics[width=0.84\linewidth]{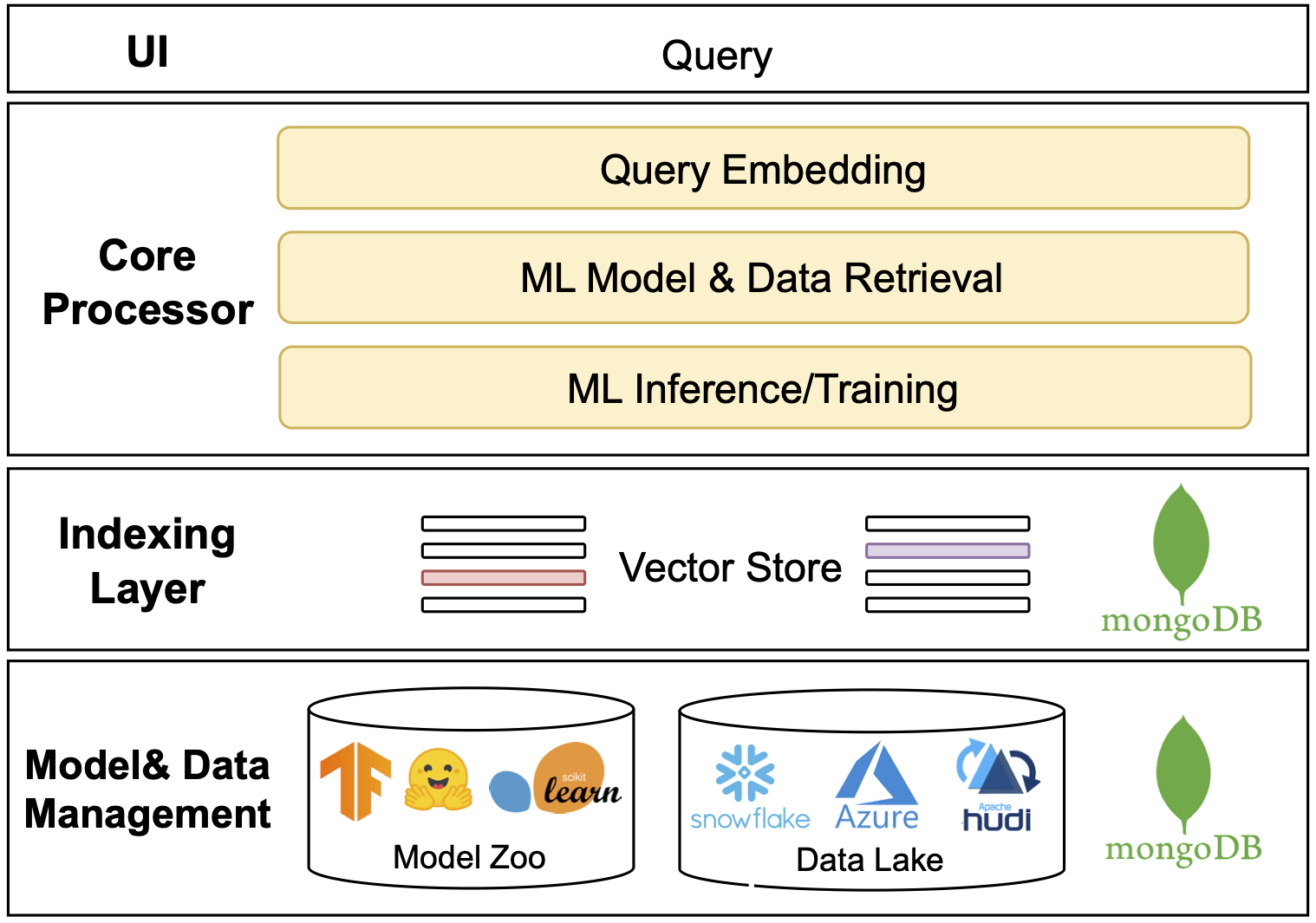}%
  \caption{Components of \texttt{LLM-PQA}}
  \label{fig:component}
   \vspace{-2mm}
\end{figure}

\texttt{LLM-PQA} consists of the following components, as shown in \autoref{fig:component}, arranged in layers that collectively contribute to answering a query.

\para{Prediction query} This is the input provided by the user, which consists of a task that requires ML model inference to obtain the results. The query is expressed in natural language. 

\para{Core processor}
The main functionalities are performed in this layer.
The layer is responsible of retrieving matched model and dataset given a user query, and then perform model inference.
We perform a vector search to retrieve the model and dataset.
The processes include i) vectorizing/indexing the query, ii) retrieving corresponding dataset and model with similarity search given the query vector, and iii) performing ML model inference to obtain the results. 
Detailed explanations are illustrated in Section~\ref{ssec:workflow}.

\para{Indexing layer}
To support vector search, datasets and models are indexed in the form of vectors. 
All the vectors, including the ones for query, datasets and models, should apply the same encoding method, using the same encoding model to generate the vectors.
The indexed entities are stored in a database system.

\para{Model, data, and metadata management}
To provide sources for the datasets and model collection, in the foundation layer of \texttt{LLM-PQA}, we employ our previous data lake \cite{hai2016constance, 10412203} and model zoo ~\cite{li2023metadata}. 
A model zoo~\cite{kaggle,onnx,li2023metadata} is a collection of diverse ML models of different capabilities, which can be updated and expanded with new models being trained.
A data lake~\cite{nargesian2019data, hai2023data} is a data management system that stores a vast amount of raw data in its native format with metadata regarding its structure and other information. 
In \texttt{LLM-PQA}, all the models and datasets are associated with a unique profile, which can be regarded as a model/dataset card.
Currently, the system includes 52 model cards and 50 dataset profiles.

\subsection{Workflow}
\label{ssec:workflow}

\begin{figure}
  \centering
  \includegraphics[width=0.8\linewidth]{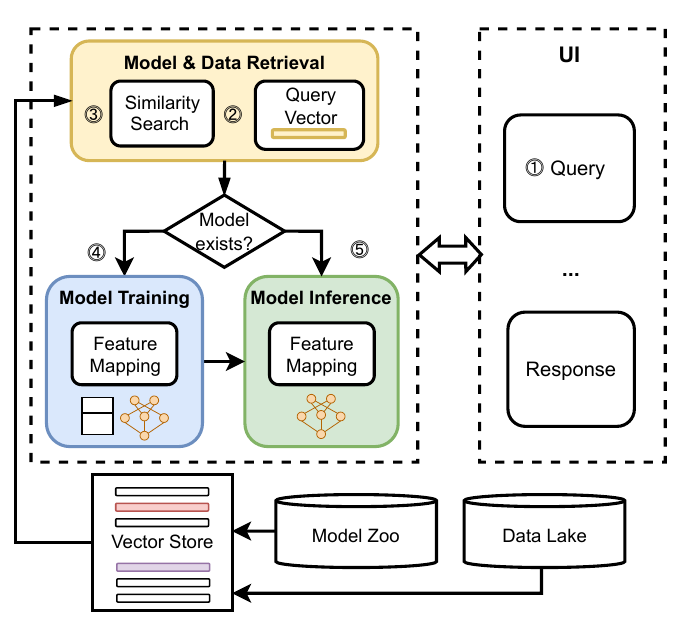}
  \caption{The workflow of answering prediction query}
  \label{fig:workflow}
   \vspace{-2mm}
\end{figure}

We present the workflow of \texttt{LLM-PQA}  in \autoref{fig:workflow}.
The depicted workflow illustrates the process of addressing the ML inference query in natural language.
The process initiates with a user query in step \ding{172}. 
The query could contain analytic requests on top of the stored datasets, encompassing predictive and classification tasks.

\begin{figure}
  \centering
  \includegraphics[width=0.88\linewidth]{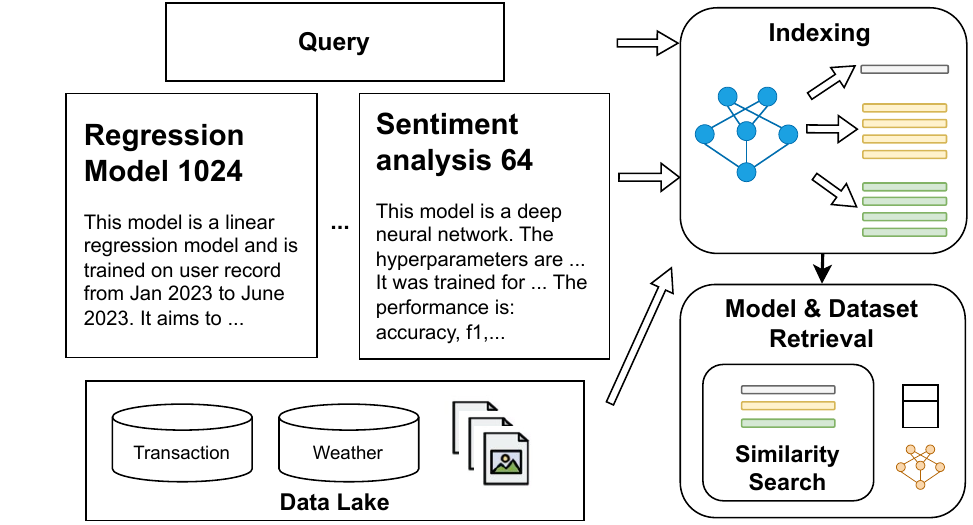}
  \caption{Retrieving model and dataset with vector search}
  \label{fig:vector}
\end{figure}

\para{Indexing query, models and datasets}
\label{ssec:index}
As the preparation process, we first index these entities into vectors, as in Figure~\ref{fig:vector}.
Each model and dataset consist of its raw files (e.g., script, weights, data documents) as well as a profile describing detailed information, e.g., metadata, statistics, etc.
We use a text encoder, `text-embedding-ada-002', from OpenAI, to generate the embedding vectors from the descriptive profiles.
The model profile includes training details, for example the training dataset, used features, performance.
The dataset profile contains information regarding, for example, the domain and features.
The vectors of models and datasets are stored in a vector store, allowing for semantically relevant retrieval.

\para{Retrieving Model/Dataset given query}
The backend takes the user query as input and first, as step \ding{173}, embeds the query into a vector, in the same way as indexing the models and datasets, as presented in Figure~\ref{fig:vector}.
With the vectorized entities, (i.e., models, datasets, and query), we perform a vector search in step \ding{174}. 
We use cosine similarity for similarity search, returning the model profile with the most similar vector. We then retrieve the model scripts based on the information in the returned profile.
We conducted a preliminary experiment on query construction. The text encoder effectively handle grammatical errors and synonyms in queries, facilitating accurate vector retrieval.
With the retrieved model, a user can verify its applicability for the request. Depending on the  feedback, the subsequent process may involve either performing only model inference, or first training then inference.

\para{Feature mapping for model inference and training}
Before executing the model, one important step is to identify the features to be used.
When there is no model available, we need to distinguish the feature columns against the label column for preparing a training dataset.
If a model is matched for the query, we identify the feature values indicated in the query and feed them to the model, using the prompt as below.

\begin{mdframed}
\begin{verbatim}
    Given the columns {self.columns} in a dataset
and a user's query related to regression analysis,
the user's query '{self.query}', the task
is to predict a numerical outcome based on
various input features. Please suggest the most 
appropriate column names for: 
1. Input variables: columns that will serve as input 
features for predicting the outcome. 
2. Output variable: the single column that represents 
the target outcome to predict. 
Other format requests ...
\end{verbatim}
\end{mdframed}

\begin{figure}
  \centering
  \includegraphics[width=0.95\linewidth]{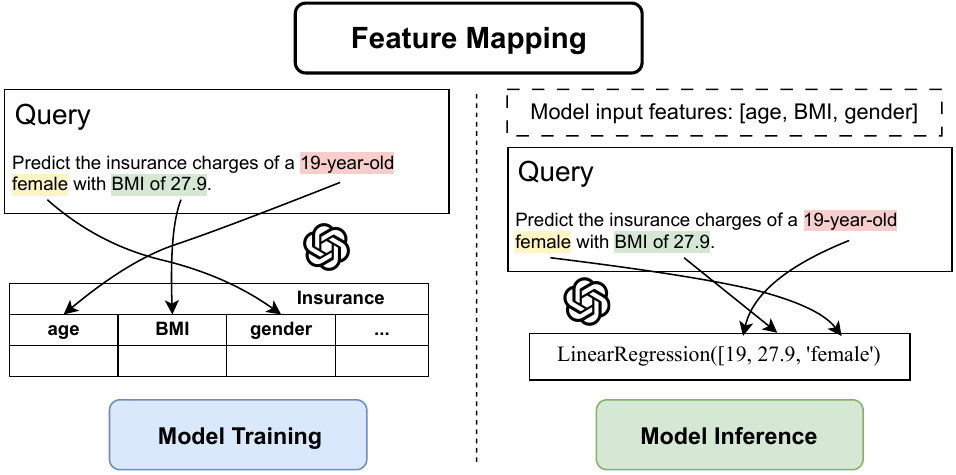}
  \caption{Identifying feature (values) from the query}
  \label{fig:features}
\end{figure}

Figure~\ref{fig:features} presents the process of identifying features for these two scenarios, i.e., model training and model inference. We use an example prediction query of predicting the insurance charges of a 19-year-old female, non-smoker, with BMI of 27.9. 
When training a new model, \texttt{LLM-PQA} sends the column names of the dataset along with the user's query to the LLM. The LLM identifies input features and output label column, such as suggesting `age, BMI, gender' for input features and `insurance charges' as the target label, as in the left panel in Figure~\ref{fig:features}. The feature columns used for training will be recorded along with the trained model.
An example prompt is shown above for identifying the features used for model training.

 
For model inference, the LLM extracts relevant feature values given the user query and model input information, i.e., types and order of the features, as in the right panel of Figure~\ref{fig:features}. For instance, from the same query, the LLM identifies the values, `19, 27.9, female', for the corresponding input features, i.e., age, BMI, and gender.

\para{Model inference}
If a model is verified to match the query, model inference is performed in this stage (step \ding{176}).
The value of different features is identified during feature mapping process, as in Figure~\ref{fig:features}.
With the feature values being fed to the model, results will be returned to the user. 
We performed various end-to-end tasks with model inference, achieving average times of 6.5 seconds for regression and 12.6 seconds for classification tasks.

\para{Model training and profiling}
There are also situations when no model is identified to answer the task, and thus model training is necessary.
A user needs to choose an ML algorithm to train, e.g., decision tree.
The dataset will be preprocessed for training with features and labels identified, as indicated previously in feature mapping step.
With the features and model algorithm selected, we train a model in an online mode.
After it is trained, a profile will be generated automatically and stored in the database, and can be further used during retrieval.
In the profile, certain information will be recorded, e.g., training dataset, features, type of algorithm, performance, etc.
We outline the structure of a model profile below.
 
\begin{mdframed}
\begin{verbatim}
Model Name: customerproductrecommender7172 
Dataset Name: trx_data
Model Overview: Aiming for a recommendation task, ...
Intended Use: ...
Technical Details:
  - Algorithm Type: ...
Model Performance:
  - Accuracy: 0.807
Limitations: ...
\end{verbatim}
\end{mdframed}

%% file: sections/demo.tex
\section{Demonstration}

\begin{figure}
  \centering
  \includegraphics[width=0.9\linewidth]{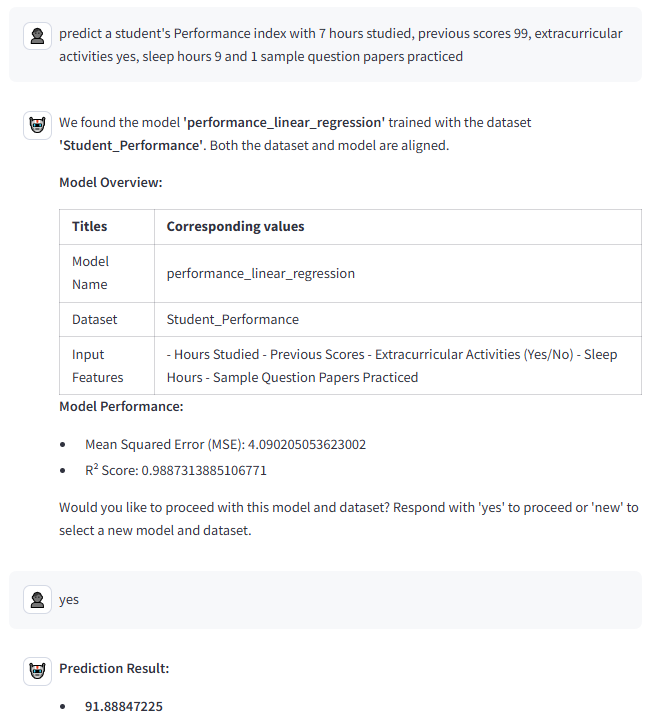}
  \caption{Interface: query answering with matched model}
  \label{fig:UI}
\end{figure}

\begin{figure}
  \centering
  \includegraphics[width=0.95\linewidth]{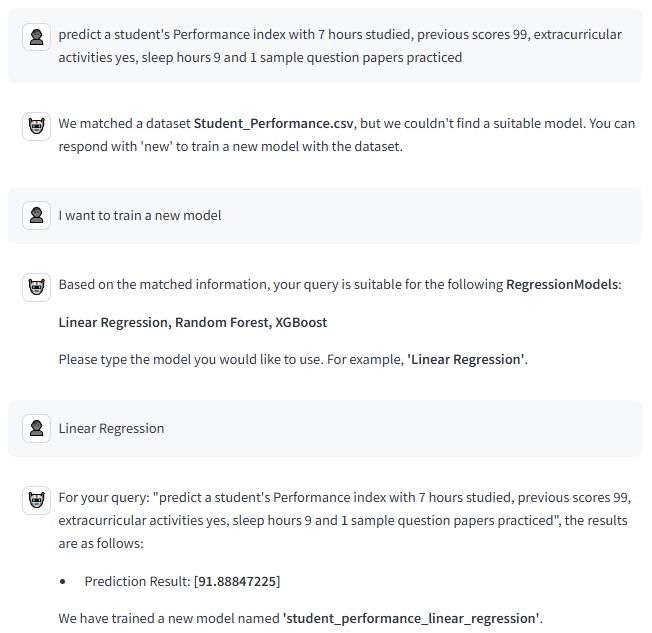}
  \caption{Interface: model training when no model is matched}
  \label{fig:UI-train}
\end{figure}

\texttt{LLM-PQA} is developed in Python and utilizes the \textit{langchain} library\footnote{\url{https://python.langchain.com/v0.2/docs/introduction/}} for tasks such as encoding entities and interfacing with the LLM API. Data storage, including model and dataset files as well as index vectors, is handled by MongoDB Atlas\footnote{\url{https://www.mongodb.com/products/platform/atlas-database}}. The user interface is designed as a chatbox, enabling users to interact with \texttt{LLM-PQA} by submitting queries and providing any required intermediate inputs. This interaction is processed by the backend, with results subsequently displayed in the interface. 
In this demo, we enable to address two types of prediction tasks, i.e., regression and classification (binary classification, and multi-label classification for recommendation).
In Figure~\ref{fig:UI} and Figure~\ref{fig:UI-train}, we showcase a scenario for regression task where a user would like to predict a student's performance.
The features provided are: studied 7 hours, previous scores of 99, with extra-curricular activities, 9 hours of sleep and practiced 1 sample question paper.

\para{Model inference with matched model}
\label{ssec:model-inference}
In Figure~\ref{fig:UI}, \texttt{LLM-PQA} first takes the user query as input, and encodes it into a vector, with the same encoding model vectorizing the model and dataset profiles, as explained in Section~\ref{ssec:index}.
A model (`performance\_linear\_regression') and a dataset (i.e., `Student\_Performance') are returned by retrieving the most similar vectors compared to the query vector.
A brief model profile is then presented as a response.
The description includes training details and performance of the model.
Once the user verify and confirm the combination of model and dataset pair, \texttt{LLM-PQA} will perform model inference. The predicted result for this query is 91.89.

\para{Online model training}
In Figure~\ref{fig:UI-train}, we showcase a scenario where no model is matched to answer the query.
As seen in the figure, \texttt{LLM-PQA} cannot provide information on the ML model. No model is identified to answer the query, while a dataset is matched for the task.
Subsequently, the tool would suggest the user to train a model given the matched dataset.
Then, the user can specify the type of model to be trained.
For each type of task, e.g., regression, there is a default type of model that will be recommended.

Afterwards, as described above, feature identification is performed to select the features and labels from the dataset.
When a model being trained, the weight file and profile are generated and stored, ready for retrieval in the future.
With the trained model, model inference is applied and the result is returned.

%% file: sections/conclusion.tex
\section{CONCLUSION AND FUTURE WORK}
In this work, we propose \texttt{LLM-PQA}, which facilitates prediction query answering in natural language. 
The vector search mechanism, matching model with given query vector, ensures that the model prediction is both precise and relevant to the query's requirements.
By integrating a data lake and model zoo, \texttt{LLM-PQA} provides access to a vast array of heterogeneous data and ML models, enhancing its capability to answer queries from a broad spectrum. 
%

For future work, we will conduct more exploration with the \texttt{LLM-PQA} framework. To enhance retrieval results, one future direction is to improve the exploitation of dataset and model information, such as using statistical and histogram data. Our recent work, \texttt{TransferGraph}~\cite{li2024model}, has shown that these relationships can be informative for predicting model performance on specific tasks. 
Moreover, future work can explore optimizing the design of entity profiles or representations to better leverage their intrinsic properties.
Additionally, further research can incorporate advanced data discovery techniques to enhance dataset searches.